%Paper: hep-th/9207058
%From: PANDAS@itsictp.bitnet
%Date: Thu, 16 Jul 92 14:16 N

\magnification=1200
\font\titlea=cmb10 scaled\magstep1

\baselineskip=12pt
\rightline{IC-92-145}
\rightline{hep-th/9207058}
\baselineskip=18pt
\vskip .5cm
\centerline{\titlea Remarks on the Additional Symmetries and W-constraints}
\centerline{\titlea in the Generalized KdV Hierarchy}
\vskip 2cm
\baselineskip=14pt
\centerline{Sudhakar Panda$\,^\ast$}
\bigskip
\centerline{and}
\bigskip
\centerline{Shibaji Roy}
\bigskip
\centerline{\it International Centre for Theoretical Physics}
\centerline{\it Trieste - Italy}
\vskip 1.0cm
\baselineskip=18pt
\centerline{\titlea Abstract}
\bigskip
Additional symmetries of the $p$-reduced KP hierarchy are generated by the
Lax operator $L$ and another operator $M$, satisfying $res~(M^n~L^{m+n/p})$ = 0
for $1~\leq~n~\leq~p-1$ and $m~\geq~-1$ with the condition that ${\partial L
\over {\partial t_{kp}}}$ = 0, $k$ = 1, 2,..... We show explicitly that the
generators of these additional symmetries satisfy a closed and consistent
W-algebra only when we impose the extra condition that ${\partial M \over
{\partial t_{kp}}}~=~0$.
\vfill
\hbox to 3.5cm {\hrulefill}\par
\item{($\ast$)} Address after 1st November 1992: Institute for Theoretical
Physics, Groningen University, 9747 AG Groningen, The Netherlands.
\eject
It is a remarkable observation that the integrable models share an interesting
relationship with the matrix model formulation of 2D quantum gravity [1-5],
2D topological gravity [6-8] and the intersection theory on the moduli space of
Riemann surfaces [9-11]. In the matrix model formulation of the 2D gravity, one
employs the method of orthogonal polynomials and makes use of the operators
$Q$ and $P$ which correspond to the insertions of the spectral parameter and
a derivative with respect to this in the matrix integral [12,13]. Since these
two operators are conjugate to each other, they satisfy the so-called \lq\lq
string
equation" [13] $[P, Q]~=~1$. As Douglas argued, these operators can be realized
in terms of some finite order differential operators and can be recognized as
the Lax-pair of an associated integrable model [14-16].

It is also well known that the formulation of Hermitian one [17] (two) [18]
matrix
model in terms of the continuum Schwinger Dyson equations gives rise to a
semi-infinite Virasoro ($W^{(3)}$) constraints on the square root of the
partition
function. The square root of the partition function is recognized as the
$\tau$-
function of the KdV (Boussinesq) hierarchy with the flow parameters being
identified with the coupling constants of the scaling operators. One of the
Virasoro constraints arising in this way is also seen to be equivalent with the
integrated form of Douglas' string equation. Further, the conjecture [17] that
all other
Virasoro and W-constraints can be obtained from the string equation itself is
also proved [19-21] in different indirect ways. In fact, a lot of efforts
has been
made in understanding the origin and the geometry of the string
equation [22,23].
It has been observed that the solution space of the string equation has
interesting connection with the Sato-Grassmannian [24,25].

The origin of the Virasoro and W-constraints can also be attributed
to the existence
of additional symmetries of the integrable hierarchy as propounded
in [26]. For
$p$-reduced KP hierarchy [27,28], the Lax operator is given by
$$
L~=~\partial^p + u_{p-2} (x,t) \partial^{p-2} +........+ u_0 (x,t)
\eqno(1)
$$
where $\partial \equiv {\partial \over {\partial x}}$ and $u_0,....,u_{p-2}$
are coefficient functions of the Lax operator depending
on the infinite number of evolution paramaters $( x, t_2, t_3,......)~\equiv
( x, t )$. The coefficient functions satisfy the following evolution equations
$$
{\partial L \over {\partial t_n}}~=~[ L^{n/p}_+~,~L ]~~~~~~~~~~n~=~1, 2,
3.......
\eqno(2)
$$
where $L^{n/p}_+$ denotes the non-negative differential part of the $n$-th
power
of the formal pseudo-differential operator $L^{1/p}$. For $n = 1$, the above
equation gives a consistency relation if we identify $t_1$ with $x$. It
is clear
that for $n = 0 (mod~p)$, the commutator in (2) vanishes and thus,
$$
{\partial L \over {\partial t_{kp}}}~=~0 ~~~~~~~~~k~=~1, 2, 3........
\eqno(3)
$$
The additional symmetries [29] of $L$ mean that $L$ does not depend on some
additional symmetry parameters $t_{mp+n, n}$. In other words,
$$
{\partial L \over {\partial t_{mp+n, n}}}\equiv [ L, ( M^n L^{m+n/p} )_- ]~=~0
\eqno(4)
$$
for $m\geq - 1$, $1 \leq n \leq p - 1$ and these flows commute with the usual
$p$-reduced KP hierarchy flows given in (2). But they do not commute among
themselves [29]. Here, the operator $M$ is defined as
$$
M~=~{1\over {p}}~K~( \sum_{n=1}^{\infty} n t_n \partial^{n-1} )~K^{-1}
\eqno(5)
$$
where $K$ is the pseudo-differential operator,
$$
K~=~1 + \sum_{i=1}^{\infty} a_i (x,t)~\partial^{-i}
\eqno(6)
$$
and known as the Zakharov-Shabat dressing operator. It satisfies the relation
$L = K \partial^p K^{-1}$ which fixes the coefficients $a_i (x,t)$ in terms
of $u_i$ and their derivatives.

It has been shown in [20,21] that the equations (3), (4) along with the
definition
of $M$ as in (5) give rise to an infinite set of constraints on the
$\tau$-function
of the $p$-reduced KP hierarchy which are claimed to be the
semi-infinite Virasoro
and W-constraints appearing in the matrix models, as discussed earlier.
However,
that is not quite true, as we explicitly show for the case of $p = 3$ that the
constraints generated in this way do not satisfy a closed and consistent
$W^{(3)}$
algebra [30]. Further, we show that the situation can be salvaged if we impose
the extra condition that $M$ satisfies
$$
{\partial M \over {\partial t_{kp}}}~=~0~~~~~~~~~~~k = 1, 2, 3,.........
\eqno(7)
$$
Only then, the generated constraints satisfy a consistent W-algebra and also
they
exactly match with the matrix model results in [17,18]. We would also like to
point out here that this extra condition is not necessary to be imposed to
obtain the right Virasoro constraints.

We now specialize for the case of $p = 3$, i.e. we are dealing with the
3-reduced
KP (Boussinesq) hierarchy. The Lax operator for this hierarchy is
$$
L~=~\partial^3~+~4 u \partial ~+~(2 u^{\prime} + v )
\eqno(8)
$$
where $u^{\prime} \equiv {\partial u\over \partial x}$ and the particular
form of
the Lax operator is chosen such that it transforms covariantly under the
conformal transformation [31]. The Douglas' string equation, for the two-matrix
model, interpolating between various critical points is given as,
$$
\sum_{k=1\atop i=1,2}^\infty ( k + {i\over 3}) t_{3k+i}~ [ L, L^{{3(k-1)+i}
\over 3}_+]~ = ~1
\eqno(9)
$$
where we have introduced an infinite number of evolution parameter $t_{3k+i}$
proportional to $-1/(k+i/3)$ and we recall that $L$ satisfies ${\partial L
\over {\partial t_{3k}}} = 0$. The operator $M$ in this case has the form
$$
M~=~{1\over 3}~K (\sum_{n=1}^\infty n t_n \partial^{n-1} ) K^{-1}
\eqno(10)
$$
The operators $L$ and $M$ satisfy the following commutation relations :
$$
\eqalignno{&[~L^{{1\over 3}}, M~]~=~{1\over 3} &(11 a)\cr
           &[~L, M L^{-{2\over 3}}~]~=~1 &(11 b)\cr
           &[~M, L^{-{2\over 3}}~]~=~{2\over 3}~L^{-1} &(11 c)\cr}
$$
By explicitly writing $M L^{-{2\over 3}}$ with the form of $M$ as given in
(10),
it is a simple exercise to check that the string equation (9) is equivalent to
$$
[~L, ( M L^{-{2\over 3}})_+~]~=~1
\eqno(12)
$$
Also, from (11), (12) and using the residual symmetry of the Zakharov-Shabat
dressing operator $K$ [21] it can be seen that
$$
( M L^{-{2\over 3}} )_-~=~{1\over 3}~L^{-1}
\eqno(13)
$$
Equation (13), in turn, implies that for $n \geq 0$
$$
( M L^{n+{1\over 3}} )_-~=~( ( M L^{-{2\over 3}} )_- L^{n+1} )_-~=~0
\eqno(14)
$$
In terms of the additional symmetries, (13) and (14) simply mean that the Lax
operator is independent of the symmetry parameters $t_{3n+1, 1}$ for
$n\geq -1$.
As a consequence of (13), it is straight forward to check that $M$ and $L$
obey the following relations [20]
$$
( M^2 L^{-{4\over 3}}~- {4\over 3} M L^{-{5\over 3}}~+ {4\over 9}
L^{-2} )_-~=~0
\eqno(15a)
$$
and for $n\geq -1$
$$
( M^2 L^{n+{1\over 3}} )_-~=~0
\eqno(15b)
$$
Again, equation (15) implies that $L$ does not depend on the additional
symmetry
parameters $t_{-4,2}$, $t_{-5,1}$ and $t_{3n+1,2}$ for $n\geq -1$. It is clear
that (13) is consistent with (12). In fact, one can work out the residue (the
coefficient of the $\partial^{-1}$ term) of (13) explicitly and demonstrate
that
it is eqivalent to (9).

By making use of the vertex operator techniques of KP hierarchy [27,28], it is
shown in [20,21] that the residues of (13) and (14) reduce to the following
differential operator constraints on the $\tau$-function of the Boussinesq
hierarchy:
$$
L_n~\tau ~=~0  ~~~~~~~~n\geq -1
\eqno(16)
$$
where the operators $L_n$ (with $t_2 \equiv y$) are defined as
$$
\eqalignno{&L_{-1}~ =~\sum_{k=1\atop i=1, 2}^\infty (k + {i\over 3}) t_{3k+i}
{\partial\over \partial t_{3(k-1)+i}}~+~{2\over 3}~ x y &(17a)\cr
&L_0~=~\sum_{k=0\atop i=1, 2}^\infty (k + {i\over 3}) t_{3k+i} {\partial\over
\partial t_{3k+i}}~+~{1\over 9} &(17b)\cr}
$$
and for $n\geq 1$
$$
L_n~=~\sum_{k=0\atop i=1, 2}^\infty (k + {i\over 3}) t_{3k+i} {\partial \over
\partial t_{3k+i}}~+~{1\over 6} \sum_{i+j=3n \atop i, j \neq 0 (mod 3)}
{\partial\over \partial t_i}~{\partial\over \partial t_j}
\eqno(17c)
$$
It can be checked that for $n\geq -1$, the operators $L_n$, as given above,
satisfy the centerless Virasoro algebra [ $L_m$, $L_n$ ] = $(m~-~n)~L_{m+n}$.
It is worth noting that in the vertex operator techniques, how (13) and (14)
give rise to the above constraints is not at all transparent. However, by just
using the Lax operator approach, one can show in a more direct way that this
is indeed true. The details of this approach will be reported elsewhere [32].
Also, note that the constant term ${1\over 9}$ in the definition of $L_0$,
which
originates as an integration constant, is fixed so, such that $L_n$'s satisfy
the
Virasoro algebra. This point is not mentioned in [20].

In a similar way, one can work out the consequences of (15a) and (15b) by
considering
the residue term. After an integration with respect to $x$ and multiplication
by $\tau$, equation (15) reduces to the following differential constraints on
$\tau$ [20,21]
$$
W_n^{(3)}~\tau ~=~0
\eqno(18)
$$
for $n\geq -2$ and where $W_n^{(3)}$ is defined as
$$
\eqalign{W_n^{(3)}~=~&{1\over 27}~\sum_{i+j+k=-3n} i j k t_i t_j
t_k ~+~{1\over 9}~\sum_{i
+j -k =-3n} i j t_i t_j {\partial \over \partial t_k}\cr~&
+~{1\over 9}~\sum_{i-j-k=-3n} i t_i {\partial^2 \over {\partial t_j
\partial t_k}}~+~{1\over 27}~\sum_{i
+j+k=3n} {\partial^3 \over {\partial t_i \partial t_j \partial t_k}}\cr}
\eqno(19)
$$
It is claimed in [20,21] that the differential operators $L_n$ for $n\geq -1$
and $W_m^{(3)}$ for $m\geq -2$ as defined in (17) and (19) respectively, form
a closed $W^{(3)}$ algebra. At this stage, the following fact should be noted.
Since the $\tau$-function of the Boussinesq hierarchy satisfies [27]
$$
{\partial\over \partial x} {\partial \over \partial t_n}
\log\tau~=~res (L^{{n\over 3}})
\eqno(20)
$$
and the right hand side of this vanishes when $n=3k$ for any $k=1,2,...$, so
the $\tau$-function would not depend on these coordinates. Therefore, it
is clear
that the non-zero contributions to $W_n^{(3)}$ comes from the first term
and only from $k\neq$ 0 (mod 3),
$j,~k\neq$ 0 (mod 3) and $i,~j,~k\neq$ 0 (mod 3) respectively for the last
three
terms of (19). Keeping this fact in mind, we can compute the commutators
between
$L_n$ and $W_m$. We find, for example, that
$$
[ L_{-1}, W_{-1}^{(3)} ]~=~-~W_{-2}^{(3)}~-~{24\over 27}~ t_6
\eqno(21)
$$
where $W_{-2}^{(3)}$ has the form as given in (19). Therefore, the form of this
$W_{-2}^{(3)}$ is inconsistent with the algebra
$$
[ L_m, W_n^{(3)} ]~=~(2m -n)~W_{m+n}^{(3)}
\eqno(22)
$$
Let us also think of modifying $W_{-2}^{(3)}$ by adding to it a term
$-{24\over 27}
t_6$. This is possible because $W_{-2}^{(3)}$ has the same scaling
dimension as $t_6$
and the expression for $W_{-2}^{(3)}$ is obtained from (15a) after an
integration
with respect to $x$. So $t_6$ term can appear as an integration constant. We
mention here that $W_{-2}^{(3)}$ contains a term propertional to $t_2^3$
which infact
appears as an integration constant. Thus we write,
$$
\tilde{W}_{-2}^{(3)}~=~W_{-2}^{(3)}~-~{24\over 27}~t_6
\eqno(23)
$$
However, it does not serve our purpose since for example, we find
$$
[ L_0, \tilde{W}_{-2}^{(3)} ]~=~2~\tilde{W}_{-2}^{(3)}~+~{48\over 27}~t_6
\eqno(24)
$$
which is again inconsistent with (22). Thus, from the above analysis, it is
clear that $W_n^{(3)}$ can not be a function of $t_{3k}$ for $k$ =1, 2,......
Therefore, in all the terms in (19), $i,j,k$ should not take values
0 (mod 3). Note that
this is consistent with the matrix model result [17,18]. Upon imposing this
extra condition, equation (19) takes the form
$$
\eqalign{W_n^{(3)}~&=~{1\over 27} \sum_{i+j+k=-3n\atop {i,j,k\neq 0 (mod 3)}}
i j k t_i t_j t_k ~+~{1\over 9} \sum_{i+j-k=-3n\atop {i,j,k\neq 0(mod 3)}}
i j t_i
t_j {\partial \over \partial t_k}\cr
&+~{1\over 9} \sum_{i-j-k=-3n\atop {i,j,k\neq 0 (mod 3)}} i t_i
{\partial^2 \over
{\partial t_j \partial t_k}}~+~{1\over 27} \sum_{i+j+k=3n\atop {i,j,k\neq
0(mod 3)}} {\partial^3\over
{\partial t_i\partial t_j\partial t_k}}\cr}
\eqno(25)
$$
We can now check that (17) and (25) do indeed satisfy a consistent $W^{(3)}$
algebra:
$$
\eqalign{[ L_m, L_n ]&=~(m-n)~L_{m+n}\cr
         [ L_m, W_n ]&=~(2m-n)~W_{m+n}\cr
         [ W_m, W_n ]&=~(m-n)~\{{3\over2} (m^2 +4 m n + n^2)~+~{27\over 2}
         (m+n)
{}~+21\}~L_{m+n}\cr &-~9 (m-n) U_{m+n}~-~{1\over 10}~
\delta_{m,-n} m (m^2-1) (m^2-4)\cr}
\eqno(26)
$$
where we have defined $U_m~=~\sum_{k\leq -2} L_k L_{m-k}~+
{}~\sum_{k\geq -1} L_{m-k}
L_k$.

Let us note that the extra conditions that $i, j, k \neq 0$ (mod 3) in (25)
can not be obtained if they are obtained from (15) with the definition of $M$
as given in (10). In order to obtain (25) as a direct consequence of the
additional symmetries, we can modify the operator $M$ itself and take instead
$$
M~=~{1\over 3} K~(\sum_{n=1\atop n\neq 0 (mod 3)}^\infty n t_n
\partial^{n-1} )~K^{-1}
\eqno(27)
$$
With the above definition, we would of course have
$$
{\partial M\over \partial t_{3k}}~=~0~~~~~~~~~~~~k=1,2,3.......
\eqno(28)
$$
and the equations (11) are still valid. Therefore, the string equation (12) and
the consequences of it namely (13-15) also remain to be true. But since $M$
does
not depend on the parameters $t_{3k}$, these parameters will never appear
either
in the Virasoro or in the $W_n^{(3)}$ constraints and we have the required form
of these constraints. Note that even if we do not modify the operator $M$ to
the above form, one still obtains the correct form for the Virasoro constraints
but not for the $W$-constraints as we saw. This feature remains valid for
higher
KdV hierarchy also. Thus, for $p$-th KdV hierarchy, we need to take the
operator
$M$ as
$$
M~=~{1\over p} K~(\sum_{n=1\atop n\neq 0 (mod p)}^\infty n t_n
\partial^{n-1} )~K^{-1}
\eqno(29)
$$
in order to obtain the right $W^{(p)}$ constraints.

In conclusion, we have shown that the origin of the Virasoro and
$W$-constraints in
the $p$-reduced KP hierarchy can be attributed to the additional symmetries
associated with such an integrable system. The generators of these additional
symmetries have to be modified with appropriate changes in the operator $M$
appearing in the generators to obtain a closed and consistent $W^{(p)}$
algebra.
\vskip 1cm
{\titlea Acknowledgments}
\bigskip
We would like to thank Prof. E. Gava, Prof. K. S. Narain for clarification
on a point in [18] and Prof. A. Salam, IAEA, UNESCO and ICTP, Trieste,
for support.
\vfil\eject
{\titlea References}
\bigskip
\item{1.} D. Gross and A. Migdal, Phys. Rev. Lett. 64 (1990), 127; Nucl. Phys.
B340 (1990), 333.
\item{2.} M. Douglas and S. Shenkar, Nucl. Phys. B335 (1990), 635.
\item{3.} E. Brezin and V. Kazakov, Phys. Lett. B236 (1990), 144.
\item{4.} E. J. Martinec, Univ. of Chicago preprint EFI-90-67, (1990).
\item{5.} L. Alvarez-Gaume, CERN preprint CERN-TH-6123/91 (1991).
\item{6.} R. Dijkgraaf and E. Witten, Nucl. Phys. B342 (1990), 486.
\item{7.} E. Verlinde and H. Verlinde, Nucl. Phys. B348 (1991), 457.
\item{8.} J. Distler, Nucl. Phys. B342 (1990), 523.
\item{9.} E. Witten, Surveys In Diff. Geom. 1 (1991), 243; preprint
IASSNS-HEP-91/74.
\item{10.} R. Myers and V. Periwal, Nucl. Phys. B333 (1990), 536.
\item{11.} M. Kontsevich, Max-Planck Institute preprint MPI/91-47; MPI/91-77.
\item{12.} S. Chadha, G.Mahoux and M. L. Mehta, J. Phys. A14 (1981), 579.
\item{13.} M. R. Douglas, Phys. Lett. B238 (1990), 176.
\item{14.} P. D. Lax, Comm. Pure Appl. Math. 21 (1968), 467; ibid 28
(1975), 141.
\item{15.} L. D. Fadeev and L. A. Takhtajan, \lq\lq Hamiltonian methods in the
Theory of Solitons", Springer-Verlag (1987).
\item{16.} A. Das, \lq\lq Integrable Models", World Scientific (1989).
\item{17.} M. Fukuma, H. Kawai and R. Nakayama, Int. Jour. Mod.
Phys. A6 (1991), 1385.
\item{18.} E. Gava and K. S. Narain, Phys. Lett. B263 (1991), 213.
\item{19.} R. Dijkgraaf, E. Verlinde and H. Verlinde, Nucl. Phys. B348
(1991), 435.
\item{20.} J. Goeree, Nucl. Phys. B358 (1991), 737.
\item{21.} M. Adler and P. Van-Moerbeke, Brandeis preprint (1992).
\item{22.} G. Moore, Comm. Math. Phys. 133 (1990), 261.
\item{23.} V. Kac and A. Schwartz, Phys. Lett. B257 (1991), 329.
\item{24.} M. Fukuma, H. Kawai and R. Nakayama, Comm. Math. Phys. 143
(1992), 371.
\item{25.} K. N. Anagnostopoulos, M. J. Bowick and A. Schwarz, Syracuse
preprint
SU-4238-497 (1991).
\item{26.} A. Y. Orlov and E. I. Shulman, Lett. Math. Phys. 12 (1986), 171.
\item{27.} E. Date, M. Kashiwara, M. Jimbo and T. Miwa in \lq\lq Nonlinear
Integrable Systems", eds. M. Jimbo and T. Miwa, World Scientific (1983).
\item{28.} G. Segal and G. Wilson, Publ. Math. IHES 63 (1985), 1.
\item{29.} L. A. Dickey, Univ. of Oklahoma preprint (1992).
\item{30.} A. B. Zamolodchikov, Theo. Math. Phys. 65 (1985), 1205.
\item{31.} A. Das and S. Roy, Int. Jour. Mod. Phys. A6 (1991), 1429.
\item{32.} S. Panda and S. Roy in preparation.
\bye